# Chern insulators and high Curie temperature Dirac half-metal in two-dimensional metal-organic frameworks

Cui-Qun Chen,[1] Xiao-Sheng Ni,[1] Dao-Xin Yao,[1, *] and Yusheng Hou [1, †]

**AFFILIATIONS**

[1] Guangdong Provincial Key Laboratory of Magnetoelectric Physics and Devices, State Key Laboratory of Optoelectronic Materials and Technologies, Center for Neutron Science and Technology, School of Physics, Sun Yat-Sen University, Guangzhou, 510275, China

**ABSTRACT**

Two-dimensional (2D) magnetic materials with nontrivial topological states have drawn considerable attention recently. Among them, 2D metal-organic frameworks (MOFs) are standing out due to their advantages, such as the easy synthesis in practice and less sensitivity to oxidation that are distinctly different from inorganic materials. By means of density-functional theory calculations, we systematically investigate the electronic and topological properties of a class of 2D MOFs $X(C_{21}H_{15}N_3)$ (X = transition metal element from 3$d$ to 5$d$). Excitingly, we find that $X(C_{21}H_{15}N_3)$ (X = Ti, Zr, Ag, Au) are Chern insulators with sizable band gaps (~7.1 meV). By studying a four-band effective model, it is revealed that the Chern insulator phase in $X(C_{21}H_{15}N_3)$ (X = Ti, Zr, Ag, Au) is caused cooperatively by the band inversion of the $p$ orbitals of the $C_{21}H_{15}N_3$ molecule and the intrinsic ferromagnetism of $X(C_{21}H_{15}N_3)$. Additionally, $Mn(C_{21}H_{15}N_3)$ is a Dirac half-metal ferromagnet with a high Curie temperature up to 156 K. Our work demonstrates that 2D MOFs $X(C_{21}H_{15}N_3)$ are good platforms for realizing Quantum anomalous Hall effect and designing spintronic devices based on half-metals with high-speed and long-distance spin transport.

Authors to whom correspondence should be addressed:
[Yusheng Hou, houysh@mail.sysu.edu.cn; Dao-Xin Yao, yaodaox@mail.sysu.edu.cn]

Two-dimensional (2D) materials have been an emerging and rapidly growing research field recently, as they manifest many exotic phenomena, such as quantum spin Hall (QSH) states and quantum anomalous Hall effect (QAHE). [1] Thanks to the reduced dimensionality, 2D materials exhibit quantum confinement effects, less quenched orbital magnetic moments and stronger spin magnetic moments which are not found in tree-dimensional materials. [2-6] Of particular interest are 2D metal-organic frameworks (MOFs) consisting of metal atoms and organic molecules. 2D MOFs are standing out for their combined advantages of both organic and inorganic materials, such as the easy synthesis in practice, low cost, insensitivity to oxidation, etc. [7, 8] To date, 2D MOFs with nontrivially topological states have attracted a great deal of attention. [9] In 2013, Wang *et al*. predicted that triphenyl-Metal has topologically nontrivial phases with a nonzero $Z_2$ invariant ($Z_2 = 1$) that are robust against lattice strain. [10] In the following years, electron-doped $Ni_3(C_{18}H_{12}N_6)_2$ [11] and Au-DCB [12] were reported to display QSH phases. Besides, Cu-DCA [13] and $Tl_2Ph_3$ [14] are shown to be topological insulators with spin Chern number $C_S$= -1 and the flat Chern band with an extremely large band gap is found in the latter. In 2018, Zhang *et al*. found non-Dirac band dispersions in TPyB-Co.[15] On the other hand, the emergent QAHE draws considerable interest in MOFs with time reversal symmetry broken. In 2013, Wang *et al*. theoretically demonstrated that $Ni_3C_{12}S_{12}$ can realize nontrivial topological states with nonzero Chern number ($C_N = \pm 1$).[16] After that, $Mn_2C_{18}H_{12}$,[17] 2D indium-phenylene organometallic framework (IPOF) $In_2(C_6H_4)_3$ and Bi-DCB [12] were all predicted to manifest QAHE. In triangular and bi-triangular lattices, Zhang *et al*. theoretically revealed that TPyB-Ta and TPyB-Hf [18] experience spin-orbit coupling (SOC)-induced topological phase transition from a topologically trivial phase to a topologically nontrivial one. Until now, the realization of QAHE has been experimentally achieved in a few of materials [19-22] although the QAHE has been theoretically proposed in many 2D systems. Hence, it is of significance to find more 2D topological materials which should be easily synthesized.

On the other hand, 2D FM Dirac half-metals (DHMs) are very promising for developing next-generation spintronic devices, as they carry two intriguing properties simultaneously, i.e., the fully spin polarization and massless Dirac fermions. Furthermore, they may overcome the challenges in spintronics, such as injection, long-distance transport of spin-polarized carriers, as well as spin manipulation and so on. [23, 24] DHMs can be regarded as a special type of half-metals where one of spin channels has a Dirac cone whereas the other one is gapped. In practical technologies, half-metals show tremendous potential for the ideal electrodes in spintronic devices. [25] Half-metal electrodes can not only provide fully spin-polarized currents and large magnetoresistance (MR) in giant MR and tunneling MR devices, but can also realize efficient spin injections into semiconductors, thereby opening the way for more applications from low-power spin transistors to spin-based quantum computing. [26, 27] Although 2D ferromagnetic (FM) materials have been widely sought,[28-30] intrinsic 2D FM materials especially with high Curie temperatures and full spin polarizations are still under urgent need currently.

In this Letter, inspired by the experimental synthetizations of $Fe(C_{21}H_{15}N_3)$ and $Cu(C_{21}H_{15}N_3)$,[31,32] we performed a systematic first-principles study on the electronic and magnetic properties of a class of 2D triangular lattice MOFs $X(C_{21}H_{15}N_3)$ (X = transition metal elements from $3d$ to $5d$). We find that $X(C_{21}H_{15}N_3)$ (X=Ti, Zr, Ag, Au) are metallic with their valence and conduction bands touching at Γ point. Interestingly, they display M-shaped valence bands or W-shaped conduction bands near Γ point when SOC is considered. By calculating their edge states, we confirm that they are magnetic topological insulators with a Chern number $C_N=\pm 1$. The study of a four-band effective model reveals that the band inversion of the $p$ orbitals of the $C_{21}H_{15}N_3$ molecule and their intrinsic ferromagnetism cooperatively give rise to the Chern insulator phase in $X(C_{21}H_{15}N_3)$ (X = Ti, Zr, Ag, Au). Moreover, we predict that $Mn(C_{21}H_{15}N_3)$ is a FM DHM with a strong ferromagnetic Heisenberg exchange interaction up to 11.9 meV. As a result of such large Heisenberg exchange interaction, our Monte Carlo simulations show that $Mn(C_{21}H_{15}N_3)$ has a high Curie temperature (~156K). Our work shows that 2D MOFs $X(C_{21}H_{15}N_3)$ have promising potential applications in next-generation spintronic devices.

The 2D MOFs $X(C_{21}H_{15}N_3)$ (X = transition metal) have a triangular lattice structure with the symmetry of $P\bar{6}M2$ space group. As shown in FIG. 1(a), the unit cell of $X(C_{21}H_{15}N_3)$ contains one metal atom and one 1,3,5-tris(pyridyl)benzene (TPyB) molecule. If regarding the lattice structure of $X(C_{21}H_{15}N_3)$ as a quasi-honeycomb lattice, we can see that metal atoms and TPyB molecules are located at A and B sites, respectively. So far, TPyB-Fe and TPyB-Cu have already been synthesized in experiments.[31, 32] In structural relaxations, we take their experimentally measured lattice constants (13.54 Å) as the initial lattice constants for our studied $X(C_{21}H_{15}N_3)$. The optimized lattice constants of the 2D magnetic MOFs TPyB-Ti, TPyB-Mn, TPyB-Zr, TPyB-Ag and TPyB-Au are 13.60, 13.52, 13.71, 13.85 and 13.80 Å, respectively. Besides, our DFT calculations reveal that TPyB-X (X = Ti, Mn, Zr, Ag, Au) possess different net magnetic moments. More explicitly, TPyB-Mn, TPyB-Ti, TPyB-Zr, TPyB-Ag and TPyB-Au have a net magnetic moment of 3, 2, 2, 1 and 1 $\mu_B$ per unit cell, respectively. It is worth noting that the magnetic moments of the metal atoms are zero in TPyB-Ag and TPyB-Au, which implies that Ag and Au are in $Ag^{1+}$ and $Au^{1+}$ states respectively.

To investigate the magnetic properties of the above-mentioned magnetic TPyB-X and considering the large distance between the nearest neighboring (NN) transition metals, we adopt a spin Hamiltonian that consists of the NN Heisenberg exchange coupling:

$$H = -J\sum_{\langle i,j \rangle} \vec{S}_i \cdot \vec{S}_j \qquad (1)$$

In Eq.(1), $J$ is Heisenberg exchange coupling parameter and $\vec{S}_i$ is the spin at site $i$. Note that FM and antiferromagnetic (AFM) Heisenberg exchange couplings are characterized by positive and negative $J$, respectively. Our calculations show that TPyB-Ti, TPyB-Mn, TPyB-Zr, TPyB-Ag and TPyB-Au favor the FM ground state over the AFM spin order by 66.2, 47.4, 2.0, 9.8, 11.9 meV per unit cell, respectively. According to Eq.(1), the NN Heisenberg exchange coupling parameters $J$ are estimated to be 16.6, 11.9, 0.5, 2.5, 3.0 meV for TPyB-Ti, TPyB-Mn, TPyB-Zr, TPyB-Ag and TPyB-Au, respectively. It is worth noting that both TPyB-Ti and TPyB-Mn have remarkably large FM Heisenberg exchange couplings.

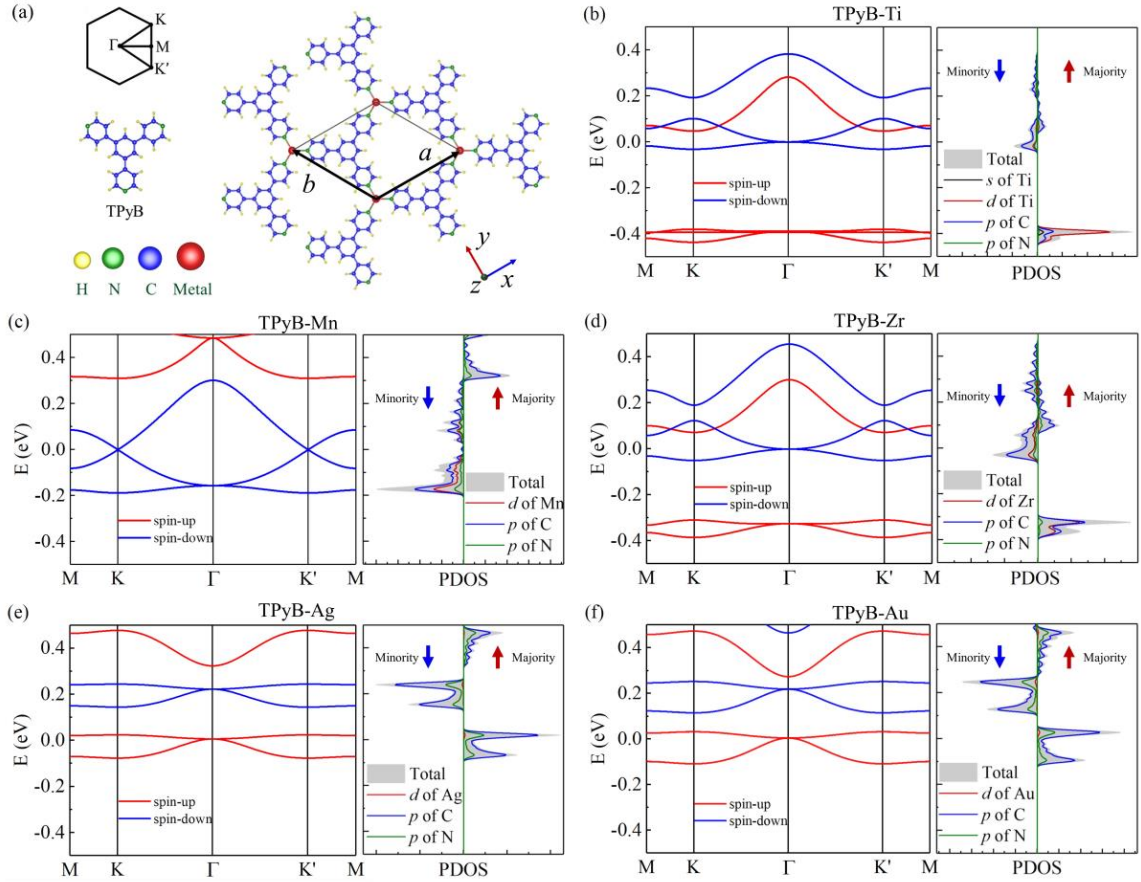

*FIG. 1. (a) Top view of the crystal structure of TPyB-X. The black lines indicate the unit cell of TPyB-X that consists of one TPyB molecule and one transition metal atom. Yellow, green, blue, red balls represent hydrogen, nitrogen, carbon and transition metal atoms, respectively. The left top inset shows the first Brillouin zone and the high-symmetry points. The left medium inset shows the TPyB molecule. The blue (red) arrow indicates the x (y) axis of the coordinate system. (b)-(f) DFT calculated spin-polarized band structures and projected density of states (PDOS) of TPyB-Ti, TPyB-Mn, TPyB-Zr, TPyB-Ag and TPyB-Au without SOC. In the left panels of (b)-(f), the red and blue curves denote spin-up (majority) and spin-down (minority) bands, respectively. In the right panels of (b)-(f), the*

*total DOS is shown by the filled grey area and the majority (minority) states are indicated by the red (blue) arrows.*

Figure 1(b)-(f) show the DFT calculated spin-polarized band structures of TPyB-X (X=Ti, Mn, Zr, Ag, Au) without SOC. Since the K and K' points could be different in systems without time-reversal symmetry, [33,34] a path of M-K-Γ-K'-M is chosen to display the band. First, one can see that all TPyB-Xs are metals, because their maximum valence bands (MVBs) and minimum conduction bands (MCBs) are touching at either Γ or K point. Interestingly, TPyB-Mn manifests a Dirac cone with linear dispersions in the vicinity of K point [FIG. 1(c)]. Second, the bands around Fermi levels are basically contributed by spin-up states in TPyB-Ti, TPyB-Zr, TPyB-Ag and TPyB-Au whereas those of TPyB-Mn are from spin-down states. Especially, the spin-up channels are gapless but the spin-down ones are gapped in TPyB-Ag and TPyB-Au whereas the situation is opposite in TPyB-Mn. Lastly, we can clearly see from the PDOS curves [the right panels in FIG. 1(b)-(f)] that the bands around Fermi levels mostly originate from the *p* orbitals of C and N atoms. Therefore, it is reasonable that a low-energy *k·p* model with *p* orbitals as the basis should capture the main physics in these systems.

Here, we study the effect of SOC on the electronic and magnetic properties of TPyB-X (X=Ti, Mn, Zr, Ag, Au). First of all, we calculate their magnetic anisotropy energy (MAE) defined as $E_{MAE} = E_x - E_z$, where $E_x$ and $E_z$ are the energies when their magnetic moments are perpendicular and parallel to the material planes, respectively. This definition means that a positive (negative) MAE indicates an out-of-plane (in-plane) magnetic easy axis. As listed in Table SI, we can see that TPyB-Mn has an in-plane magnetic easy axis while other four TPyB-X (X=Ti, Zr, Ag, Au) have an out-of-plane ones. Figure 2 shows the DFT with SOC calculated band structures of TPyB-X (X = Ti, Zr, Ag, Au) when their magnetic moments point perpendicularly. Compared with the band structures without SOC as shown in FIG. (b) and (d)-(f), we can clearly see that TPyB-X (X = Ti, Zr, Ag, Au) opens a sizable gap at Γ point. More explicitly, the gaps are 1.34, 5.81, 1.96 and 7.12 meV for TPyB-Ti, TPyB-Zr, TPyB-Ag and TPyB-Au, respectively. Interestingly, TPyB-Ti and TPyB-Zr have an observable M-shaped valence bands around Γ point whereas TPyB-Ag and TPyB-Au have obvious W-shaped conduction bands. Such features hint possible band inversions around Γ point and thereby nontrivial topological properties in TPyB-X (X = Ti, Zr, Ag, Au).

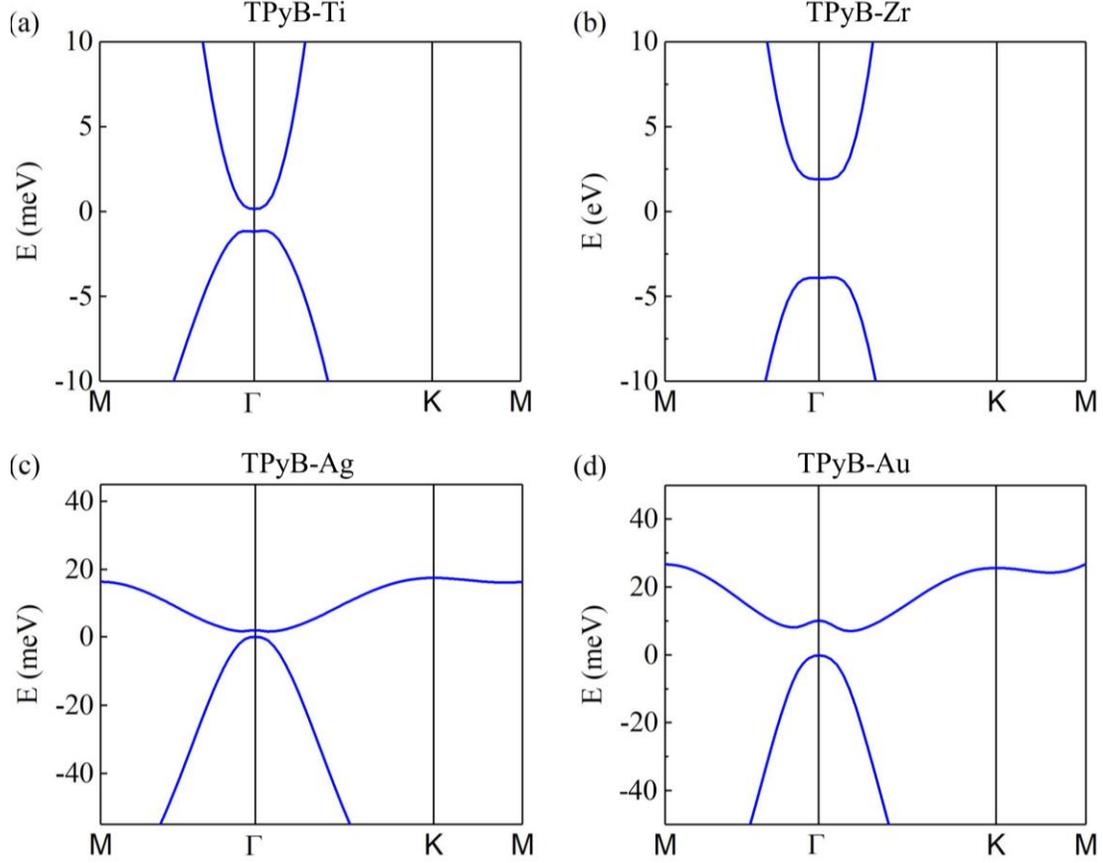

*FIG. 2. The DFT+SOC calculated band structures of (a) TPyB-Ti, (b) TPyB-Zr, (c) TPyB-Ag and (d) TPyB-Au, respectively. The Fermi levels are set to zero in (a)-(d).*

To explore the possible nontrivial topological properties in TPyB-X (X = Ti, Zr, Ag, Au), we first calculate their berry curvatures and corresponding Chern numbers. Here, we take TPyB-Au as an example as it has the largest gap among them. As shown in FIG. 3(a), the Berry curvature presents two peaks with huge values in the vicinity of Γ point. Through looking into the DFT+SOC calculated band structure, we see that the two peaks are located near the valleys of the W-shaped conduction bands. This clearly suggests that the W-shaped conduction band indeed results from the band inversion in TPyB-Au. Taking the time-reversal symmetry breaking and the band inversion into account, it is natural to expect a Chern insulator phase in TPyB-Au. To verify this, we calculate the Chern number, $C_N$, with Wannier90. As expected, the calculated $C_N$ turns out to be 1 when the Fermi level is located in the gap [FIG. 3(b)]. Accordingly, the nanoribbon of TPyB-Au has a single one-dimensional chiral edge state connecting the valance and conduction bands [FIG. 3(f)]. These unequivocally indicate that TPyB-Au is a Chern insulator and can realize QAHE. Similar to TPyB-Au, TPyB-X (X=Ti, Zr, Ag) are Chern insulators with Chern number $C_N = \pm 1$ as well, and their one-dimensional chiral edge states are distinctly displayed in FIG. 3(c)-(e).

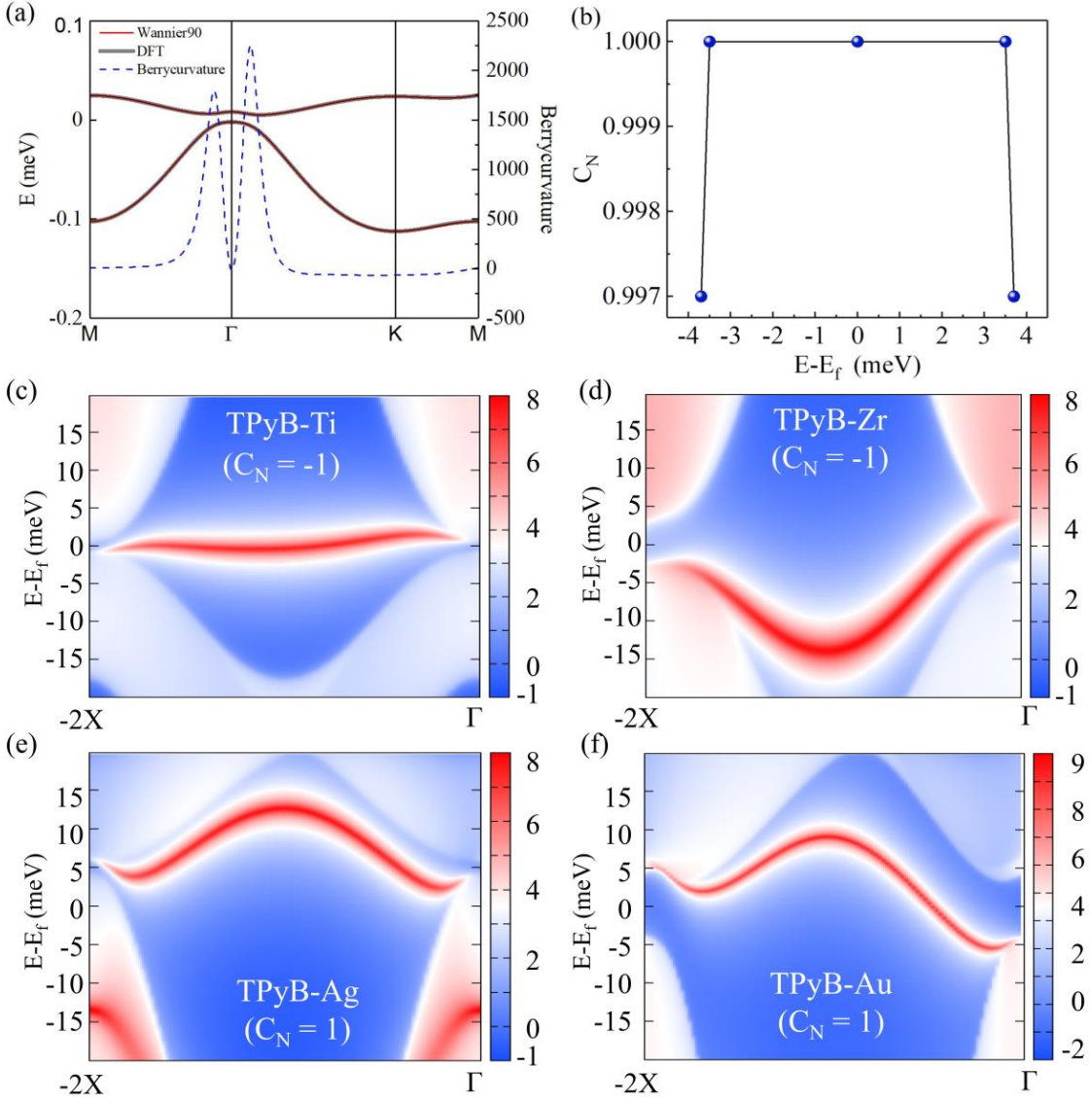

FIG. 3. Topological properties of TPyB-X (X=Ti, Zr, Ag, Au). (a) Berry curvature and DFT+SOC calculated band structure along M-Γ-K-M of TPyB-Au. (b) Dependence of Chern number on the position of fermi level in TPyB-Au. (c)-(f) show the chiral edge states of the semi-infinite TPyB-Ti, TPyB-Zr, TPyB-Ag, TPyB-Au, respectively.

To gain more insights into the nontrivial topological properties of TPyB-X (X = Ti, Zr, Ag, Au), we study a k·p model with the basis of $p_x$ and $p_y$ orbitals as the four bands near Fermi level are mainly contributed by p orbitals. For convenience, $p_x$ and $p_y$ orbitals are linearly combined to form $p_\pm = p_x \pm ip_y$. Associated with the spin-up (↑) and spin-down (↓) states, the basis is $\{|p_+,\uparrow\rangle, |p_-,\uparrow\rangle, |p_+,\downarrow\rangle, |p_-,\downarrow\rangle\}$. With the consideration of $C_3$ symmetry in TPyB-X, a four-band effective Hamiltonian can be written as follows: [9,35,36]

$$H(k) = H_0(k) + H_{soc}(k) + H_M(k)$$

$$= \begin{bmatrix} \alpha k^2 & \beta k_-^2 & 0 & 0 \\ \beta^* k_+^2 & \alpha k^2 & 0 & 0 \\ 0 & 0 & \alpha k^2 & \beta k_-^2 \\ 0 & 0 & \beta^* k_+^2 & \alpha k^2 \end{bmatrix} + \begin{bmatrix} \lambda_{soc} & 0 & 0 & 0 \\ 0 & -\lambda_{soc} & 0 & 0 \\ 0 & 0 & -\lambda_{soc} & 0 \\ 0 & 0 & 0 & \lambda_{soc} \end{bmatrix} - \begin{bmatrix} M & 0 & 0 & 0 \\ 0 & M & 0 & 0 \\ 0 & 0 & -M & 0 \\ 0 & 0 & 0 & -M \end{bmatrix} \quad (2).$$

In Eq. (2), $H_0$ is the hopping term with the related parameters $\alpha$ and $\beta$; the in-plane $k$-vector is $k_\pm = k_x \pm ik_y$; $H_{soc}$ and $H_M$ represent atomic SOC and exchange field from the intrinsic ferromagnetism of TPyB-X, respectively, and $\lambda_{soc}$ and M are their corresponding SOC and exchange field strengths, respectively. Thanks to the $C_3$ symmetry, the degenerate quadratic valence and conduction bands, whose fermi velocities are determined by $\alpha$ and $\beta$, touch each other at Γ point. Once a nonzero SOC is included in this model, a band gap opens around Γ point. Furthermore, a large enough exchange field will separate the spin-up and spin-down bands from each other. As a result, the quadratic degeneracy at Γ point vanishes and QAHE is realized. By fitting the DFT calculated bands around Γ point with this $k \cdot p$ model, we obtain the fitted relevant parameters and list them in Table SII. It is worth noting that the strong SOC-induced band gap of TPyB-Au is three times bigger than that of a similar 2D MOF TPyB-Cu which has been reported before [4].

Excitingly, we find that TPyB-Mn is a 2D high Curie temperature DHM ferromagnet. As shown in FIG. 1(c), the spin-down bands cross Fermi energy whereas the spin-up ones are gapped. This indicates that the spin-up and spin-down channels manifest as insulator and metal, respectively. Additionally, there are linearly dispersed bands in the vicinity of K and K´ points. These features characterize that TPyB-Mn is a FM DHM. In addition, the Fermi velocity of TPyB-Mn is evaluated to be $1.29 \times 10^5 \, m/s$, which is prominently advantageous for high-speed and long-distance spin transports [1]. Such large Fermi velocity can be ascribed to the delocalized *p* orbitals and small SOC strengths of C and N atoms.

Considering that the Curie temperature ($T_C$) of the FM TPyB-Mn is crucial to its practice applications in electric devices, we utilize Monte Carlo simulations to determine its $T_C$, based on the Heisenberg model [Eq.(1)] with the estimated Heisenberg exchange coupling $J = 11.9$ meV. The variation of specific heat versus temperature is displayed in FIG. 3 (a). This curve presents an obvious $\lambda$ shape with a peak locating at the temperature of 156 K, which indicates that the $T_C$ of TPyB-Mn is 156 K. Such $T_C$ of TPyB-Mn is highly attractive, as the reported $T_C$ of 2D ferromagnets are rather low, such as 80 K for $VI_3$ monolayer, 98 K for $VCl_3$ monolayer,[37] 100 K for $NiBr_3$ monolayer. [38] Hence, the high $T_C$ of TPyB-Mn makes it as a good platform for designing next-generation spintronic devices.

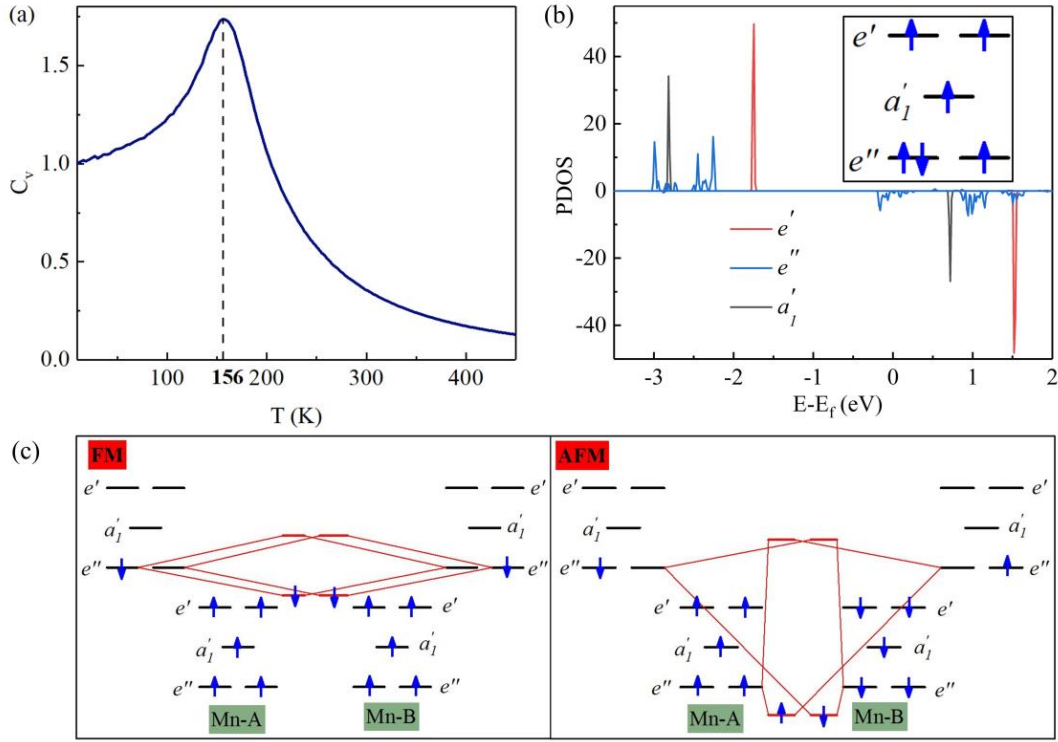

FIG. 4. (a) The dependence of specific heat, $C_v$, on temperature (T). (b) PDOS (in unit of state/eV) of Mn 3d orbitals of TPyB-Mn. The inset sketches the energy levels of Mn 3d

*orbitals. (c) Sketch of the orbital hybridizations between two magnetic $Mn^{1+}$ ions. The left and right panels are for the FM and AFM orders, respectively. The horizontal lines with/without blue arrows stand for the energy levels with occupied/unoccupied electrons. Blue arrows represent spins.*

In order to understand the underlying mechanism of the NN Heisenberg exchange coupling, we first investigate the electronic configuration of magnetic $Mn^{1+}$ ions in TPyB-Mn. By combining the Bader charge analysis and the PDOS as shown in FIG. 4(b), we find that one electron of Mn atom is released to the surrounding TPyB molecules and there are six electrons in the 3*d* orbitals of magnetic $Mn^{1+}$ ions. Due to the crystal field with the $D_{3h}$ symmetry, the originally degenerate five 3*d* orbitals are split into $a_1'$ singlet ($d_{z^2}$), $e'$ doublet ($d_{x^2-y^2}, d_{xy}$) and $e''$ doublet ($d_{xz}, d_{yz}$). From the PDOS as shown in FIG. 4(b), the energy sequences of $a_1'$ singlet, $e'$ and $e''$ doublets are sketched in the inset of FIG. 4(b). According to the Hund's rule, five electrons of $Mn^{1+}$ ions occupy the five 3*d* orbitals with spin-up states while the left one electron occupies the $e''$ doublet with spin-down states [see the inset of FIG. 4(b)]. Consequently, such occupation configuration gives rise to a magnetic moment of 4 $\mu_B$ per a Mn atom, consistent with our DFT calculations. When taking the TPyB molecule mediated hopping (parameterized by *t*) between the NN magnetic $Mn^{1+}$ pairs and the correlations (parameterized by *U*) among the 3*d* electrons of Mn into account, the energy gains are $E_{gain-FM} = -2t$ and $E_{gain-AFM} = -2t^2/U$ for the FM and AFM orders [FIG. 4(c)], respectively. Because *U* is larger than *t*, the FM order has a bigger energy gain than the AFM order. As a result, TPyB-Mn favors a FM magnetic ground state instead of an AFM one.

In conclusion, we systematically investigate the electronic and topological properties of a class of 2D MOFs $X(C_{21}H_{15}N_3)$ (X=transition metal elements from 3*d* to 5*d*). First of all, TPyB-Ti, TPyB-Zr, TPyB-Ag and TPyB-Au are demonstrated to be Chern insulators with $C_N = \pm 1$ and sizable gaps up to 7.1 meV. The nontrivially topological nature in these materials originates from the SOC-induced band inversion of the $p$ orbitals of the $C_{21}H_{15}N_3$ molecule and their intrinsic ferromagnetism. Second, we find that TPyB-Mn is 2D FM DHM metal whose spin-up and spin-down channels manifest as insulator and metal, respectively. Especially, its linearly dispersed bands in the vicinity of K and K´ points have a large Fermi velocity which is highly desirable for high-speed and long-distance spin transports. More importantly, our Monte Carlo simulations indicate that its Curie temperature is about 156 K. Our work provides theoretical guidelines for further experimental research to find topological quantum states and design spintronic devices based on 2D MOFs.

See supplementary material for the details of computational methods, magnetic and structural information as well as the related parameters of $k·p$ model.

This work was supported by NSFC-12104518, NKRDPC-2018YFA0306001, NKRDPC-2017YFA0206203, NSFC-92165204, NSFC-11974432, GBABRF-2022A1515012643, GBABRF-2019A1515011337, Shenzhen International Quantum Academy (Grant No. SIQA202102), and Leading Talent Program of Guangdong Special Projects (201626003). Calculations are performed at Tianhe-II.

## AUTHOR DECLARATIONS

**Conflict of Interest**

The author has no conflict of interest to declare.

**Authors Contributions**

**Cui-Qun Chen:** Investigation (equal); Methodology (equal); Writing – original draft (equal). **Xiao-Sheng Ni:** Writing – original draft (equal). **Dao-Xin Yao:** Investigation (equal); Funding acquisition (equal); Writing – review and editing (equal). **Yusheng Hou:** Conceptualization (equal); Funding acquisition (equal); Investigation (equal); Project administration (equal); Resources (equal); Supervision (equal); Writing – review and editing (equal).

## DATA AVAILABILITY

The data that support the findings of this study are available from the corresponding author upon reasonable request.